\documentclass{emulateapj}
\usepackage{graphicx,amsmath,amsfonts,amssymb,subfigure}
\newcommand{\kms}{\,km~s$^{-1}$}

\newcommand{\etal}{{et al.~}}

\newcommand{\kpc}{\>{\rm kpc}}

\newcommand{\Msun}{\>{\rm M_{\odot}}}
\newcommand{\Lsun}{\>{\rm L_{\odot}}}
\newcommand{\Cote}{C\^ot\'e}
\def\Sec{${}^{\prime\prime}$\llap{.}}

\begin{document}
\submitted{Accepted to ApJ, 2011 October 7}

\title{Structure and Dynamics of the Globular Cluster Palomar 13\altaffilmark{*}}

\author{J.~D.\ Bradford\altaffilmark{1,2}}
\author{M.\ Geha\altaffilmark{2}, R.~R.\ Mu\~noz\altaffilmark{2,3}, F.~A.\ Santana\altaffilmark{2,3}}
\author{J.~D.\ Simon\altaffilmark{4}}
\author{P.\ \Cote \altaffilmark{5}, P.~B.\ Stetson\altaffilmark{5}}
\author{E.\ Kirby\altaffilmark{6,7}}
\author{S.~G.~Djorgovski\altaffilmark{6,8}}

\altaffiltext{*}{The data presented herein were obtained at the
  W.M. Keck Observatory, which is operated as a scientific partnership
  among the California Institute of Technology, the University of
  California and the National Aeronautics and Space
  Administration. The Observatory was made possible by the generous
  financial support of the W.M. Keck Foundation.}

\altaffiltext{1}{Department of Physics, Central Connecticut State University, 1615 Stanley Street, 
New Britain, CT 06050. jeremydbradford@gmail.com}

\altaffiltext{2}{Astronomy
  Department, Yale University, New Haven, CT~06520.
  marla.geha@yale.edu}

\altaffiltext{3}{Departamento de Astronom\'ia, Universidad de Chile,
  Casilla 36-D, Santiago, Chile}

\altaffiltext{4}{The Observatories of the Carnegie Institution of Washington,
   813 Santa Barbara Street, Pasadena, CA~91101}

\altaffiltext{5}{National Research Council of Canada, Herzberg Institute of
  Astrophysics, 5071 West Saanich Road, Victoria, BC V9E 2E7, Canada}

\altaffiltext{6}{California Institute of Technology, Department of
  Astronomy, MS 249-17, Pasadena, CA  91106}

\altaffiltext{7}{Hubble Fellow}

\altaffiltext{8}{Distinguished Visiting Professor, King Abdulaziz University, Jeddah, Saudi Arabia}

\begin{abstract}
\renewcommand{\thefootnote}{\fnsymbol{footnote}}

We present Keck/DEIMOS spectroscopy and CFHT/MegaCam photometry for
the Milky Way globular cluster Palomar~13.  We triple the number of
spectroscopically confirmed members, including many repeat velocity
measurements.  Palomar~13 is the only known globular cluster with
possible evidence for dark matter, based on a Keck/HIRES 21 star
velocity dispersion of $\sigma=2.2\pm 0.4$\kms.  We reproduce this
measurement, but demonstrate that it is inflated by unresolved binary
stars.  For our sample of 61 stars, the velocity dispersion is
$\sigma=0.7^{+0.6}_{-0.5}$\kms.  Combining our DEIMOS data with
literature values, our final velocity dispersion is
$\sigma=0.4^{+0.4}_{-0.3}$\kms.  We determine a spectroscopic
metallicity of [Fe/H]= $-1.6\pm0.1$ dex, placing a 1-sigma upper limit of
$\sigma_{\rm[Fe/H]}\sim$ 0.2\,dex on any internal metallicity spread.
We determine Palomar~13's total luminosity to be $M_V=-2.8\pm0.4$,
making it among the least luminous known globular clusters.  The
photometric isophotes are regular out to the half-light radius and
mildly irregular outside this radius.  The outer surface brightness
profile slope is shallower than typical globular clusters
($\Sigma\propto~r^{\eta},\eta=-2.8\pm 0.3$).  Thus at large radius,
tidal debris is likely affecting the appearance of Palomar~13.
Combining our luminosity with the intrinsic velocity dispersion, we
find a dynamical mass of of $M_{1/2}=1.3^{+2.7}_{-1.3}\times
10^3M_{\odot}$ and a mass-to-light ratio of
$M/L_V=2.4^{+5.0}_{-2.4}M_{\odot}/L_{\odot}$.  Within our measurement
errors, the mass-to-light ratio agrees with the theoretical
predictions for a single stellar population.  We conclude that, while
there is some evidence for tidal stripping at large radius, the
dynamical mass of Palomar~13 is consistent with its stellar mass and
neither significant dark matter, nor extreme tidal heating, is
required to explain the cluster dynamics.

\end{abstract}

\keywords{globular clusters: individual (Palomar 13) -- galaxies:
  kinematics and dynamics --dark matter}

\section{Introduction}

Globular clusters are among the oldest structures in the universe.
Current models favor globular cluster formation via the collapse of
cold gas within a larger galactic environment \citep{fall85a,
  kravstov05a, muratov10a}.  Recently, improved evidence for internal
abundance ratios of elements heavier than hydrogen in some of the most
luminous globular clusters \citep[e.g.,][]{lee09a,cohen10a,milone11a} has renewed
interest in the idea that these objects might instead have formed at
the center of their own dark matter halos \citep{peebles84a} or are
the stripped nuclei of dwarf galaxies \citep{bekki03a}, thus providing
a deeper gravitational potential for the retention of enriched
material.  Observational evidence, however, argues against dark matter
in globular clusters, at least at the present time.  Photometric
studies of cluster profiles \citep{conroy10a} and dynamical studies of
Milky Way globular clusters \citep{baumgardt09a, lane10a, hankey10a}
show no evidence for dark matter, at least out to the tidal radius in
these systems.

\begin{figure*}[t!]
\epsscale{0.7}
\plotone{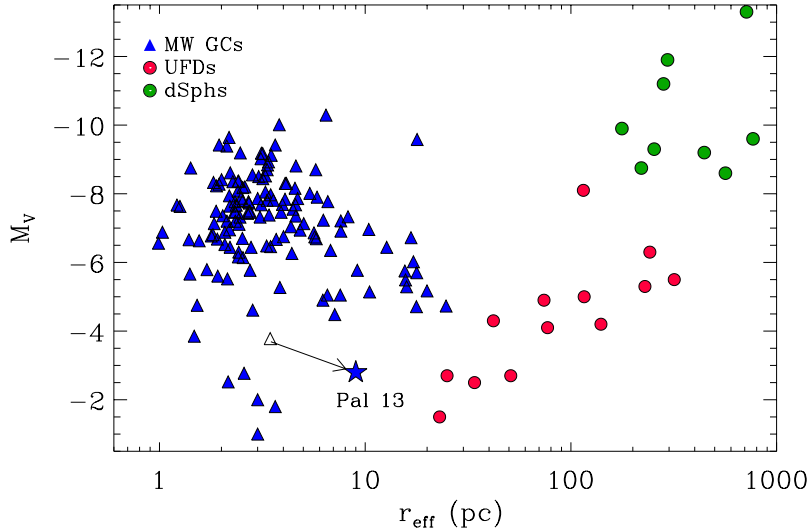}
\caption{Absolute magnitude versus half-light radius for all Milky Way
globular clusters (blue triangles, data from Harris~2010), dSphs (green circles) and
ultra-faint galaxies (red circles).    The open triangle represents
Pal\,13 based on literature values, the blue star is
the position of Pal\,13 based on quantities we determine
based on deeper CFHT/MegaCam photometry.  \label{fig_sizeMV}}
\end{figure*}

An interesting possible exception to the conclusion that globular
clusters do not contain dark matter is Palomar~13 (Pal\,13), a low
luminosity Milky Way globular cluster \citep[$M_V \sim
-3.7$;][]{harris10a} at a Galactocentric distance of 25\,kpc.  The
combined size and luminosity of Pal\,13 (see Figure~\ref{fig_sizeMV})
is more similar to the Milky Way ultra-faint galaxies than a typical
Milky Way globular cluster \citep{martin08a}.  Motivated in part by
its low luminosity and relatively large size, \citet[][hereafter
C02]{cote02a} measured velocities for 21 member stars in Pal\,13 based
on Keck/HIRES (High Resolution Echelle Spectrometer) spectroscopy.
Their velocity dispersion of $2.2\pm 0.4$\kms\ is well in excess of
that predicted from the stellar mass of Pal\,13 alone.  In contrast,
velocity dispersions of six other extended outer halo globular
clusters measured by these authors are consistent with their stellar
masses \citep{baumgardt09a}.  C02 also measure a shallow surface
brightness profile for Pal\,13 and a King tidal radius that is
significantly larger than the predicted tidal radius based on the
stellar mass of the cluster.  C02 interpreted the inflated velocity
dispersion and shallow luminosity density profile of Pal\,13 as due
either to tidal heating during a recent perigalacticon passage,
implying that the cluster is not in dynamical equilibrium, or the
presence of dark matter.  If Pal\,13 is in dynamical equilibrium, its
measured velocity dispersion implies the highest mass-to-light ratio
known in a globular cluster, $M/L_V = 40^{+24}_{-17} \Msun/\Lsun$.

The proper motion of Pal\,13, measured by \citet{siegel01a} via
photographic plates, implies that this cluster is on a highly
eccentric orbit near perigalacticon.  This is consistent with the
interpretation that Pal\,13 is in the final stages of tidal
disruption, and its increased velocity dispersion is due to tidal
heating.  However, N-body simulations by \citet{kupper11a} are unable
to reproduce either the inflated velocity dispersion or shallow
density profile using the measured orbit.  Interestingly, these
authors can reproduce the photometric properties of Pal\,13 assuming
that instead it is near apogalacticon, i.e.,~at the farthest point in
its orbit.  In this case, the observed surface brightness profile is
not due to tidal heating, but rather due to projection effects as
unbound stars, removed in the course of slow tidal evaporation, are
compressed near apogalacticon.  In these simulations, binaries are still
required to explain the observed velocity dispersion.

A high unresolved binary star fraction is a third proposed explanation
for the large velocity dispersion observed in Pal\,13.
\citet{clark04a} estimated the Pal\,13 binary star fraction as
$30\pm4\%$ based on a secondary red main sequence and the presence of
a large number of blue straggler stars.  This is higher than observed
for more massive globular clusters, but consistent with that of lower
mass globular clusters.  \citet{blecha04a} measured a lower velocity
dispersion ($0.6-0.9\pm0.3$\kms) for Pal\,13 based on eight member
stars.  Seven of these stars overlap with C02 (note Blecha et
al.~state only six), including three stars which are identified as
velocity variables in one or both samples.  Given the small number of
stars, it is yet unclear whether binaries can indeed provide an
explanation for the high velocity dispersion in Pal\,13 without
invoking tidal heating or the presence of dark matter.

Here we present Keck/DEIMOS spectroscopy and CFHT/MegaCam imaging of
the Milky Way globular cluster Pal\,13.  The paper is organized as
follows: in \S\,\ref{sec_data} we discuss data acquisition and
reduction for both our spectroscopic and photometric datasets,
including estimates of structural parameters.  In
\S\,\ref{sec_membership} we discuss the techniques used to identify
spectroscopic members.  In \S\,\ref{sec_results} we determine the
velocity dispersion, mass and mass-to-light ratio of Pal\,13, and
compare to previous results.  In \S\,\ref{sec_results} we also
examine the blue straggler population of Pal\,13.  Finally, in
\S\,\ref{sec_discussion} we interpret our results, concluding that the
kinematics of Pal\,13 are consistent with its stellar mass and find
only mild evidence for tidal effects at large radii.

Throughout this paper, we adopt a distance to Pal~13 of 24.3\,$\kpc$
\citep[R$_{\rm GC} = 25.3$\,kpc; ][]{cote02a} and reddening of E(B-V) =
0.11 \citep{schlegel98a}.  These values are in good agreement with our
own isochrone fitting of the photometry described in
\S\,\ref{subsec_phot}.

\section{Data Analysis} \label{sec_data}

\subsection{Spectroscopic Targeting}

We select spectroscopic targets in Pal\,13 based on
Canada-France-Hawaii Telescope (CFHT) CFH12K and Keck/LRIS (Low
Resolution Imaging Spectrometer) imaging as described in C02.  We
supplement these data at large radius with wider field photometry from
P.~Stetson's online
database\footnote{See~http://cadcwww.dao.nrc.ca/community/STETSON/standards/}.
Candidate Pal\,13 member stars were identified in $V$- and $I$-bands,
and $V$- and $R$-band when $I$-band was not available.  Targets were
selected based on proximity to a \citet{girardi02a} theoretical
isochrone for an age of 12\,Gyr and a metallicity of [Fe/H]$= -1.7$
\citep{zinn82a, friel82a, cote02a}.  This isochrone was used for
target selection only; a slightly better fitting isochrone is used in
the subsequent analysis (see Figure~\ref{fig_full_cmd}).  We determine a
star's distance to the isochrone via the quantity $d = \sqrt{[(V-I)_*
  - (V-I)_{\rm iso}]^2 + (V_* -V_{\rm iso})^2}$.  Stars on the red
giant branch ($V\le$ 20.5) within d = 0.1\,mag of the isochrone were
given the highest priority for spectroscopy; stars within 0.2\,mag of
the isochrone were given lower priority.  On the subgiant branch and
main sequence ($V>$ 20.5) and on the horizontal branch the selection
region was widened to within 0.2 and 0.3 mag of the isochrone for the
higher and lower priority categories, respectively.  Candidate blue
straggler stars were chosen in a box blueward of and brighter than the
main sequence turnoff/subgiant branch region.  Following the
photometric selection, spectroscopic targeting priorities were set to
favor known radial velocity members from C02 and \citet{blecha04a},
and likely proper motion members from \citet{siegel01a}.

\subsection{Spectroscopic Observations and Data Reduction} \label{opticalprops}
%MG section

The spectroscopic data were taken with the Keck~II 10-m telescope and
the DEIMOS spectrograph \citep{faber03a}.  Eight multislit masks were
observed in Pal\,13 in September 2008, with a ninth mask obtained in
October 2009.  Exposure times, observation dates, and other observing
details are given in Table~1.  The masks were observed with the
1200~line~mm$^{-1}$\,grating covering a wavelength region
$6400-9100\mbox{\AA}$.  The spectral dispersion of this setup was
$0.33\mbox{\AA}$ pixel$^{-1}$, equivalent to R=6000 for our $0$\Sec$7$
wide slitlets, or a FWHM of $1.37\mbox{\AA}$. The spatial scale was
$0$\Sec$12$~per pixel.  The minimum slit length was $5''$ which allows
adequate sky subtraction; the minimum spatial separation between slit
ends was $0$\Sec$4$ (three pixels).

Spectra were reduced using a modified version of the {\tt spec2d}
software pipeline (version~1.1.4) developed by the DEEP2 team at the
University of California-Berkeley for that survey. A detailed
description of the reductions can be found in
\citet{simon07a}.  The final one-dimensional spectra were rebinned
into logarithmic wavelength bins with 15\,\kms\ per pixel.  Radial
velocities were measured by cross-correlating the observed science
spectra with a series of high signal-to-noise stellar templates.  We
employ the telluric absorption lines in the stellar spectra to
calculate and apply a correction for the radial-velocity errors that
arise from a mis-centering of the star in the spectrograph slit.  Each
science spectrum is cross-correlated with a hot stellar template where
the telluric absorption lines are the dominant spectral features.
Since these lines are illuminated by the stellar flux, this defines
the star's photocenter within the slit in velocity space.  We applied
both a telluric and heliocentric correction to all velocities
presented in this paper.

We determined the random component of our velocity errors using a
Monte Carlo bootstrap method.  Noise was added to each pixel in the
one-dimensional science spectrum, we then recalculated the velocity
and telluric correction for 1000 noise realizations.  Error bars are
defined as the square root of the variance in the recovered mean
velocity in the Monte Carlo simulations.  The systematic contribution
to the velocity error was determined by \citet{simon07a} to be
2.2\kms\ based on repeated independent measurements of individual
stars and we refer the reader to this paper for further details.  The
systematic error contribution is expected to be constant as the
spectrograph setup and velocity cross-correlation routines are
identical.  We added the random and systematic errors in quadrature to
arrive at the final velocity error for each science measurement.
Radial velocities were successfully measured for 482 of the 566
extracted spectra across the nine observed DEIMOS masks.  The fitted
velocities were visually inspected to ensure reliability.  We
identified 90 sources as background galaxies or QSOs, and obtained
repeat measurements for 62 stellar sources.  Our final stellar sample
consists of 392 measurements of 306 unique stars.

\subsubsection{Deep CFHT MegaCam Imaging of Palomar 13} \label{subsec_phot}

\begin{figure}[t!]
\epsscale{1.1}
\plotone{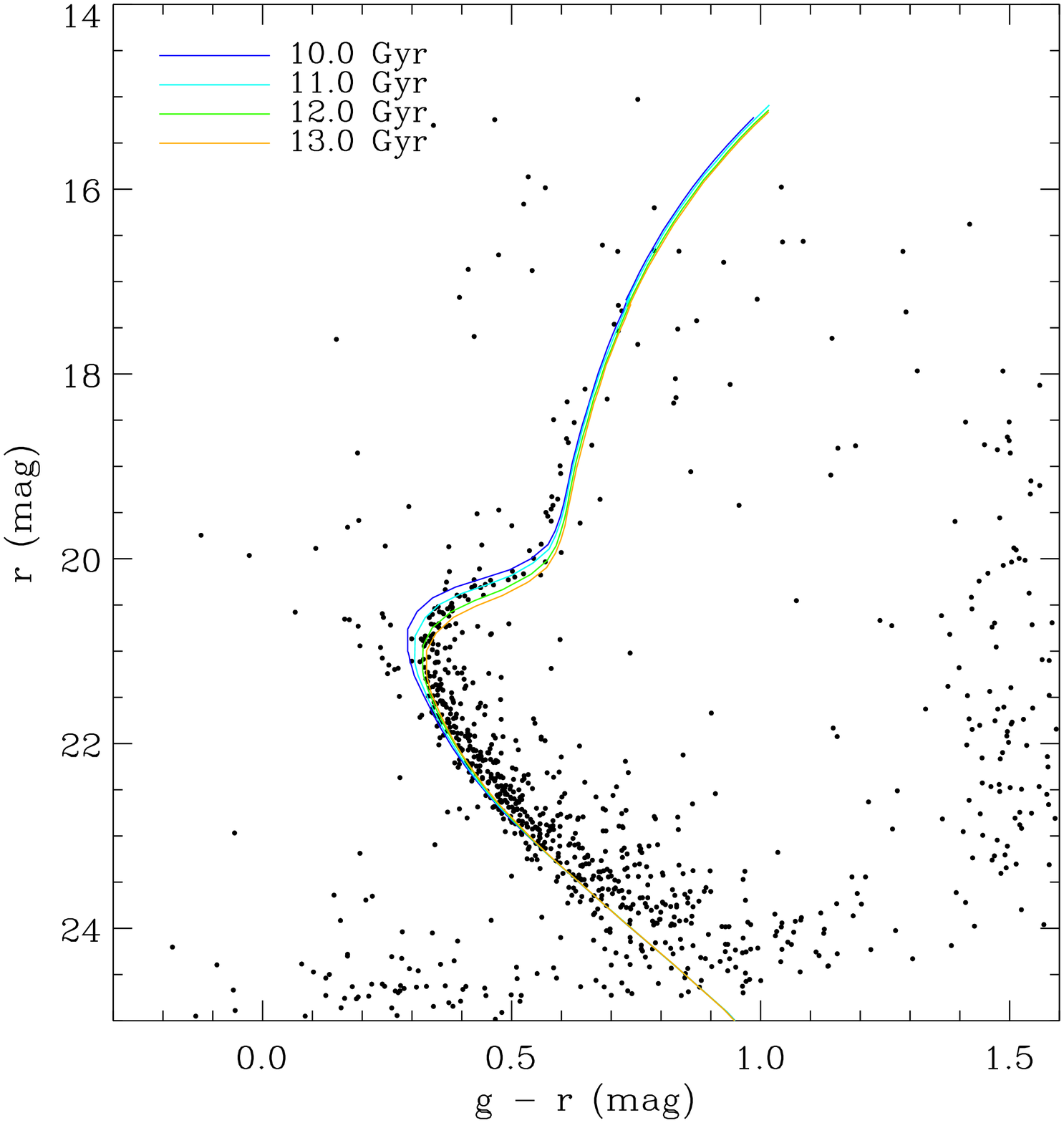}
\caption{CMD for CFHT photometric data within $5'$ of Pal\,13's
  center.  Several Girardi et al.~(2002)~isochrones are plotted with
  [Fe/H]$= -1.6$\,dex for various ages shown in the legend.\label{fig_full_cmd}}
\end{figure}

After our spectroscopic observations, we obtained significantly deeper
images of Pal\,13 using the CFHT MegaCam imager.  These images were
taken as part of a larger program aimed at obtaining deep wide-field
imaging of all bound stellar over-densities in the Galactic halo
beyond 25\,kpc (R. Mu\~noz et al.~2012, in preparation).  Pal~13 was
observed on the night of UT July 22, 2009 under dark conditions with
typical seeing of $0.7-0.9\arcsec$.  We obtained six dithered
exposures of 360\,seconds in both the $g$- and $r$-bands centered
on Pal\,13.  A standard dithering pattern was used to cover both the
small and large gaps present between chips.

Data from MegaCam were pre-processed prior to release using the
``ELIXIR" package \citep{magnier04a}. This pre-process includes bad
pixel masking, bias subtraction, flat fielding as well as
preliminary photometric and astrometric solutions.  We carried out
subsequent data reduction following the method detailed in
\citet{munoz10a} which relies on the DAOPHOT/Allstar/ALLFRAME packages
\citep{stetson94a}. To refine the astrometric solutions included in
the image headers, we have used the SCAMP package \citep{bertin06a}
using as reference the USNO-B1 catalog, which yielded
astrometric residuals of $rms\sim0.4\arcsec$.

Photometric calibration is achieved by comparing directly to Pal~13
data from the SDSS-Data Release 8 \citep[DR8;][]{dr8}.  To determine
zero points and color terms we used stars with magnitudes in the range
$18<r<21$ and $-0.25<g-r<1.0$, providing $2950$ matches between
our catalog and SDSS.  Final solutions for the zero points and color
terms yield uncertainties of $\sim0.004$\,mag and $\sim0.007$\,mag,
respectively, for both passbands.  Our photometric catalog reaches
90\% completeness at $r\sim24.1$.

\subsubsection{Improved Structural Parameters for Palomar 13}\label{subsec_struct}

\begin{figure*}[t!]
\epsscale{1.}
\plottwo{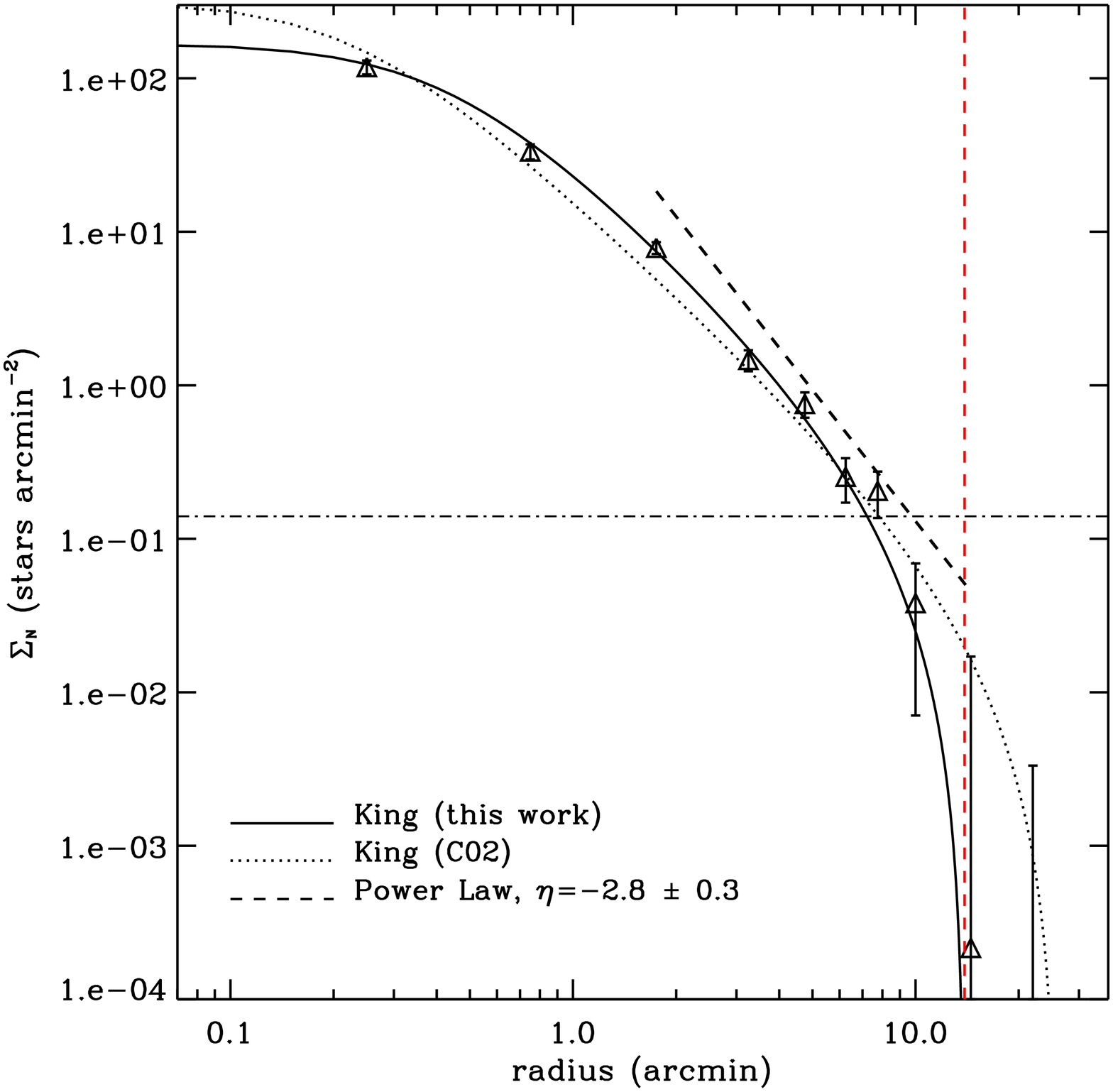}{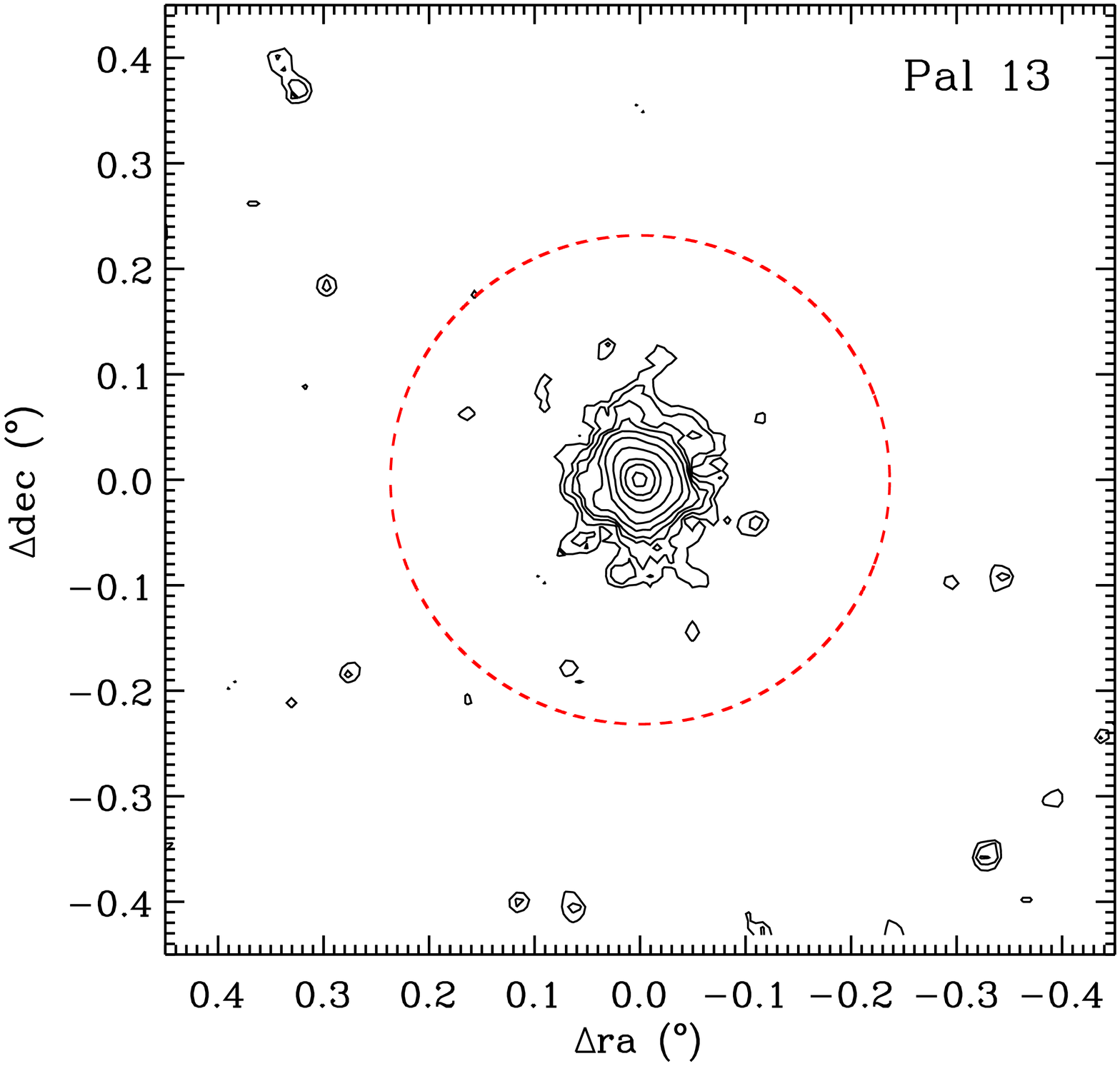}
\caption{(\textit{Left}) Pal\,13 surface brightness profile based on
  CFHT/MegaCam imaging reaching a magnitude limit of $r=24$.  Here we
  compare our best fitting King profile (solid) to that from C02
  (dotted) and parameterize the outer slope of the surface brightness
  profile (dashed).  The dot-dashed line represents the background
  level.  (\textit{Right}) The surface density profile for Pal\,13.
  The outermost contours are our $3\sigma$ confidence limits,
  corresponding to 30\,mag arcsec$^{-2}$.  The dashed red line in both
  panels is plotted at the King tidal radius of $r_t = 13.9' =
  98$\,pc.\label{fig_profile}}
\end{figure*}

Using our CFHT/MegaCam catalog, we recalculated structural parameters
for Pal\,13 using a Maximum Likelihood method as described in
\citet{munoz10a}.  In this method, a fixed analytic profile is assumed
and its free parameters are fit using all the available stars in the
CMD (color-magnitude diagram) selection window. This avoids the need
to bin or smooth data, providing more reliable estimates for systems
with low numbers of stars \citep{martin08a}.  We simultaneously
estimated the scale length, the central coordinates, the ellipticity,
the position angle and the background density of Pal\,13.  We assumed
an empirical King profile \citep{king62a}, for which we calculated a
King core ($r_{c}$) and tidal ($r_{t}$) radius.  We additionally fit a
NFW \citep{navarro97a} and a Plummer profile \citep{plummer11a}, a
good representation for dwarf galaxies, but found that these were poor
fits to the light distribution shape of Pal\,13.  The best-fitting
King parameters are $r_{c}=0.42\pm0.06\arcmin$ ($3.0\pm0.4$\,pc) and
$r_{t}=13.9\pm1.5\arcmin$ ($98\pm11$\,pc).  The two-dimensional
half-light radius of this King profile is $r_{1/2}=1.27\pm0.16\arcmin$
($9.0\pm1.1$\,pc).  The overall ellipticity of Pal~13 is consistent
with being zero and therefore its position angle is poorly
constrained.

In the left panel of Figure~\ref{fig_profile}, we show the projected
density profile for Pal\,13, overplotting our best-fit King profile.
Error bars on the data points include uncertainty in determining the
background level which dominates outside $r \sim 7'$.  We emphasize
that the King profile shown is not a direct fit to the data points,
but constructed with the best fit parameters obtained through the
maximum likelihood estimator.  We compared our best fitting King
profile to that determined in C02.  If we parameterize the outer slope
of the surface brightness profile as $\Sigma \propto r^{-\eta}$, then
typical globular clusters have slopes of $\eta \sim 4$
\citep{mclaughlin05a}.  We measure a shallow slope for Pal\,13 of
$\eta = 2.8\pm 0.3$.  This is consistent with our inferred King tidal
radius which is two times larger than the calculated tidal radius of
Pal\,13 based on its stellar mass and distance from the Milky Way (see
\S\,\ref{sec_membership}).  Our inferred slope is steeper than the C02
estimate of $\eta = 1.8\pm0.2$.  These two profiles are compared in
Figure~\ref{fig_profile}.  A key difference between the two King
profiles is the CMD region in which they were determined.  While C02
determined the profile using stars inside a box in CMD space, we have
used a window around the best-fitting isochrone.  At faint magnitudes,
there is significant contamination from unresolved galaxies, and our
CMD window should reduce this contamination. 

In the right panel of Figure~\ref{fig_profile}, we show the spatial
density map of Pal\,13.  The outer isophotes in this figure are
$3\sigma$ contours above the background, corresponding to $r\sim$30
mag arcsec$^{-2}$.  The red dashed line/circle in both panels of
Figure~\ref{fig_profile} are drawn at the King tidal radius $r_t =
13.9\arcmin$ (98\,pc). The photometric isophotes are regular out to
the half-light radius of the cluster, and appear mildly irregular
outside this radius.  This is in contrast to other faint globular
clusters such as Palomar~1 and Palomar~5 which show evidence for tidal
tails at much brighter surface brightness levels
\citep{Niederste10a,Odenkirchen03a}, although similar to that observed
for Palomar~14 \citep{sollima11a}.  We interpret the slightly
irregular isophotes and shallow density profile as mild evidence for
tidal disruption and discuss further in \S\,\ref{sec_discussion}.

Finally, we estimated the total luminosity of Pal\,13.  The published
values for Pal\,13 range between $1.2\times10^{3}$ and
$3.5\times10^{3}$\,L$_{\sun}$ \citep{siegel01a,cote02a}, making
traditional methods of adding stars' individual fluxes too sensitive
to the inclusion (or exclusion) of potential members (outliers).  To
alleviate these shot noise issues, the method used here relies solely
on the total number of stars belonging to the satellite and not on
their individual magnitudes \citep{munoz10a}.  To estimate an absolute
magnitude, we modelled the satellite's population with the
best-fitting theoretical luminosity function, in this case a $12$\,Gyr
population with [Fe/H]$=-1.6$ \citep{girardi02a}. We then integrated
the theoretical luminosity function to obtain the total flux down to a
given magnitude limit. The last step was to scale this flux using the
total number of Pal\,13 stars in our catalog down to the same
magnitude limit.  To estimate the uncertainty in the absolute
magnitude we carried out a bootstrap analysis generating $10,000$
realizations of Pal~13 from the photometric data \citep[for more
details see][]{munoz10a}.  We assumed a Chabrier initial mass function
for the estimation \citep{bruzual03b}.  We obtained
$M_{V}=-2.8\pm0.4$, or equivalently
$L_{V}=1.1^{+0.5}_{-0.3}\times10^{3}$\,L$_{\sun}$.  The central
surface brightness is $\mu_{0,V} = 23.9^{+0.6}_{-0.7}$\,mag
\,arcs$^{-2}$.  While our results are one sigma lower than published
values, it is a more robust estimate of the true luminosity of Pal\,13
as it better accounts for shot noise in this quantity.  We compare the
size and luminosity of Pal\,13 to other Milky Way stellar systems in
Figure~\ref{fig_sizeMV}.  We list our measured properties of Pal\,13
and those of C02 in Table~2 for comparison.

\begin{figure*}[t!]
\epsscale{1.1}
\plotone{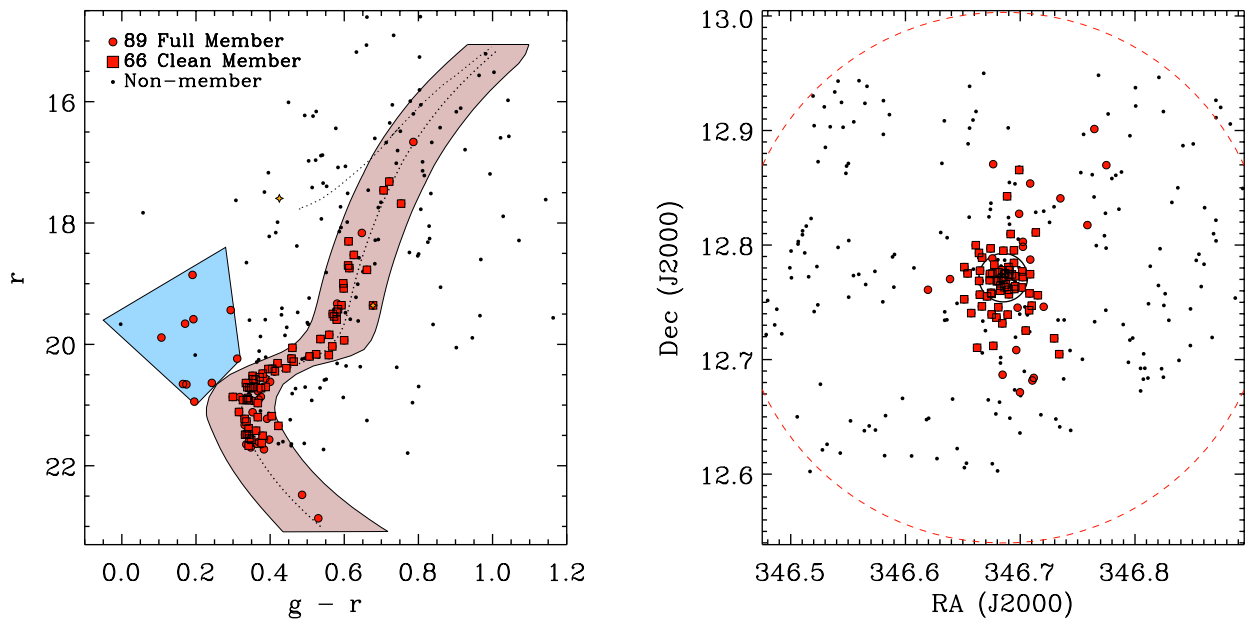}
\caption{(\textit{Left}) CMD of Pal\,13 based on CFHT imaging for all
 spectroscopic stellar targets.  The primary color selection window
  is plotted in light red with the best-fitting isochrone centered as a
  black dotted line.  The blue straggler selection window is plotted
  in light blue.  Red circles represent likely velocity members within
  the primary selection window, red squares in this region are further
  confirmed as Pal\,13 members via proper motions.  Small black dots
  represent definite non-members.  We also identify two stars from the
  C02 study as orange stars.  ORS-91 lies near the horizontal branch
  and ORS-38 lies near the RGB.  (\textit{Right}) Spatial diagram of
  Pal~13 for all spectroscopic stellar targets. Symbols are the same
  as the CMD plot on left.  The solid circle is the cluster's
  half-light radius ($r_{1/2} = 0.9'$).  The dashed red circle
  represents the King tidal radius ($r_t =
  13.9'$).  \label{fig_cmd_spat}}
\end{figure*}

\section{Membership Criteria} \label{sec_membership}

The systemic velocity of Pal\,13 lies within the velocity distribution
of foreground Milky Way stars.  We therefore selected spectroscopic
members of Pal\,13 by combining color-magnitude and velocity
criteria with proper motion information from the literature.

We first applied a color-magnitude selection window shown in
Figure~\ref{fig_cmd_spat}.  CMD selection was performed using a
12\,Gyr, [Fe/H] = $-1.6$ \citet{girardi02a} isochrone in the CFHT
photometric system, shown in Figure~\ref{fig_full_cmd} to fit Pal\,13.  The
selection window was defined as 0.1\,mag around this isochrone added
in quadrature to the 2-$\sigma$ photometric errors.  The CMD selection
window does not include blue straggler stars which are abundant in
Pal\,13 \citep{clark04a}.  We selected blue stragglers separately by
defining a generous window bluer and brighter than the main sequence
turnoff, but dimmer than the horizontal branch
(Figure~\ref{fig_cmd_spat}).  Stars in the blue straggler window are
removed from all analysis below unless specified.  We study the blue
straggler population separately in \S\,\ref{sec_blues}.

After the CMD selection, we applied a basic velocity cut to the
sample.  We chose a conservative velocity window of 10-$\sigma$ around
the systemic velocity as measured by C02.  We use the velocity
dispersion measured by C02 of $\sigma = 2.2$ \kms, thus our velocity
window corresponds to $\pm22$\kms\ around the systemic velocity of
$26.2$ \kms.  Combined with stars passing the CMD cut, this leaves 93
member candidates of Pal\,13.

Milky Way foreground stars are expected to contaminate our CMD and
velocity selected sample.  Using the Besan\c con model of the Milky
Way \citep{robin03a}, we applied the same selection criteria and
predict 5 to 7 foreground stars in our sample of 93 Pal\,13
members, depending on details of normalizing the model to data.  A
maximum of 6 of these possible 7 foreground stars are predicted in the
thin and thick disk, and 1 in the halo.  A method to identify
foreground stars is via metallicity, since the average [Fe/H] of
Pal\,13 is significantly lower than expected for the majority of
foreground Milky Way disk stars (\S\,\ref{subsec_metal}).  As
explained in \S\,\ref{subsec_metal}, we can measure
metallicities for only the 20 brightest stars in our sample.  As seen
in Figure~\ref{fig_metals}, four stars are far more metal-rich
compared to the main sample.  Three stars are at very large radius
from Pal\,13 where the contamination is greatest.  We reject these
four stars as members, leaving 89 candidate Pal\,13 stars in our
'full' sample.

\begin{figure*}[t!]
\epsscale{1.1}
\plotone{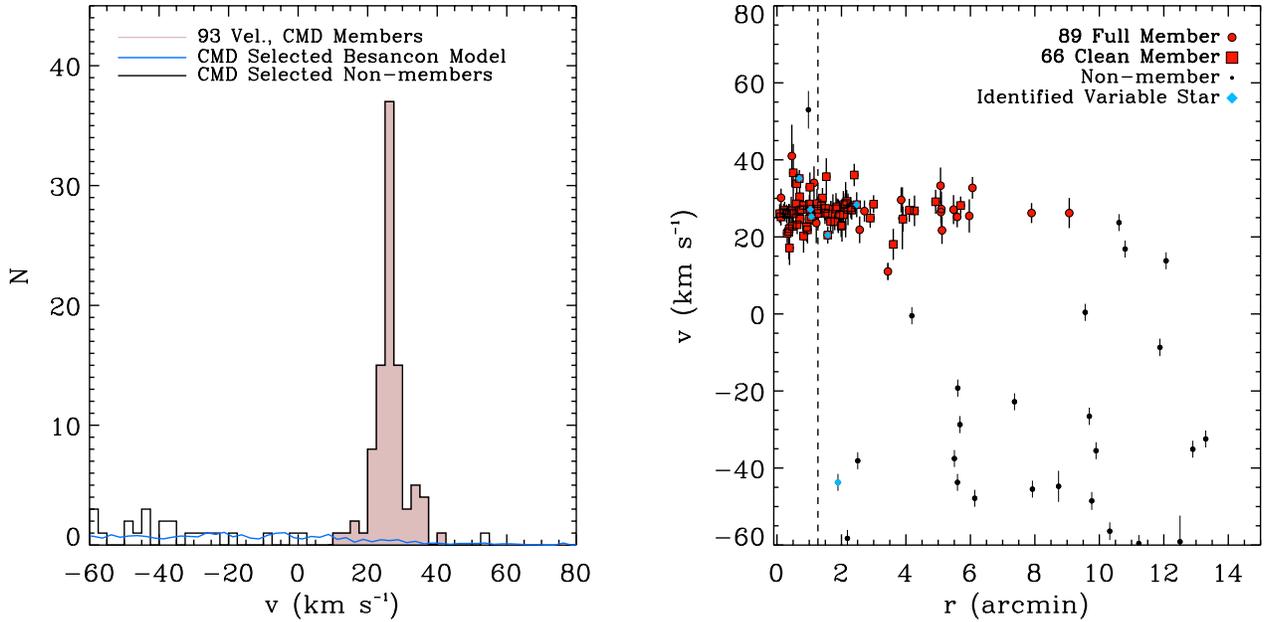}
\caption{(\textit{Left}) Velocity histogram of Pal\,13 stars
  passing our primary color-magnitude selection criterion.  We use a
  bin size of 2.5\kms\, representing the typical velocity error.  The
  normalized Milky Way model is plotted as a solid blue line and
  estimates the foreground contamination in each bin.  CMD selected
  non-members are plotted as a solid black line and excludes blue
  stragglers.  (\textit{Right}) Velocity of CMD-selected stars as a
  function of cluster radius.  The dashed line represents the
  half-light radius.  Red circles represent CMD and velocity
  selected Pal\,13 stars, red squares are stars which are further
  confirmed as Pal\,13 members via proper motions.   Small black dots
  show non-members.  We also identify binaries as blue diamonds.
  \label{fig_vel_vvsr}}
\end{figure*}

Another way to identify foreground stars is via proper motions.
Proper motions are available for 69 out of 89 candidate members (78\%)
from \citet{siegel01a}.  Three stars in this sample have large proper
motions that are inconsistent with being cluster members (assigned 0\%
membership probabilities by Siegel~et~al.), and we eliminate them from
our sample.  This leaves 66 candidate member stars of Pal\,13.  We
defined this sample of 66 stars as our 'cleaned' Pal\,13 sample, and
proceed with analysis on these stars.  We cannot be certain the
cleaned sample is free of Milky Way contamination, but estimate from
the Besan\c con models that there is likely less than one
foreground star given our sample criteria.  The results of this
selection are shown in Figure~\ref{fig_vel_vvsr}.

Tidal heating or evaporation processes will unbind stars from the
gravitational potential of Pal\,13.  These stars can be a source of
contamination if they remain in the spatial vicinity of the cluster
and would not be rejected by our methods above
\citep{klimentowski07a}.  While the three stars at largest radius
passing our CMD and velocity cuts are prime candidates for unbound
Pal\,13 stars, these were rejected based on their high metallicity and
are more likely to be Milky Way foreground stars.

To estimate at what projected radius stars are likely to be unbound
from Pal\,13, we calculate the Jacobi tidal radius assuming a
Galactocentric distance of 25.3\,kpc and a Milky Way mass inside this
radius of $3\times10^{11}\Msun$.  The Jacobi tidal radius of Pal\,13
is $6.7'$ (47\,pc), larger than the half-light radius of the cluster
$r_{\rm eff} = 1.3'$, but smaller than the inferred King tidal radius
of $r_t = 13.9'$, as noted previously by C02, \citet{siegel01a} and
\citet{kupper11a}.  Although the Jacobi radius provides only a very
approximate estimate of the tidal radius \citep{binney08a}, this
suggests that stars in the outskirts of Pal\,13 may not be bound to
the cluster.  This calculation assumes a circular orbit.  Stars in our
'clean' kinematic sample are located well inside the Jacobi tidal
radius, while our full sample contains two stars beyond this distance.

\subsection{Repeated Velocity Measurements} \label{sec_bin_pop}

We obtained numerous repeated velocity measurements of stars in
Pal\,13 in order to investigate the frequency of unresolved binary
stars and their influence on the velocity dispersion.  We measured two
or more velocities for 62 out of 306 total (36 out of 89 member)
stars.  We further fold in historical data from C02, taken ten years
prior to our DEIMOS observations, providing multiple measurements for
72 out of 306 total (42 out of 89 member) stars.  We list both DEIMOS and
HIRES velocity measurements in Table~3.  Binary fractions inferred
from our data are necessarily lower limits on the true binary fraction
and are most sensitive towards short (few day) periods, but also
include one year to ten year periods depending on the available data.
We also note that the repeated stars are heavily biased towards bright
magnitudes.

Using the error-normalized velocity difference plotted in
Figure~\ref{fig_hist_hist}, we define a star to be variable if
individual independent measurements are more than $3\sigma$ discrepant
in this quantity.  We acknowledge that the number of 'identified'
binaries depends sensitively on this choice.  We find that 5 out of 89
in our full member sample are velocity variables (6\%).  These 5 stars
are also contained in our cleaned sample.  Given the large velocity
differences (a few to tens of \kms), we assume these systems are
unresolved binary stars, as single variable stars are unlikely to be
in this region of color-magnitude space.  As expected, the fraction of
binary stars identified in our incomplete sample is far less than
$30\pm4$\% measured by \citet{clark04a}, but does provide
spectroscopic confirmation that unresolved binary stars exist in
Pal\,13.

For all stars with multiple measurements, we use the error-weighted
mean velocity in the calculations below.  Because we are measuring the
five binary stars at a random points in their orbits, it is not
guaranteed that this will be the true systemic velocity of the star.
We will therefore calculate, in \S\,\ref{sec_vel_disp}, the velocity
dispersion of Pal\,13 with and without the 5 identified binary systems.

\section{Results}\label{sec_results}

\subsection{Comparison to \Cote~\etal (2002) results}

The velocity dispersion of Pal\,13 measured by C02 of $2.2 \pm 0.4$
\kms\ is significantly larger than that predicted based on the stellar
mass alone of $\sim0.3 - 0.4$\kms.  If this velocity dispersion
accurately traces the motion of stars in a gravitational potential, it
implies over 95\% of mass in Pal\,13 is not luminous, i.e.~that
Pal\,13 has a significant dark matter component.  We measured
Keck/DEIMOS velocities for all 21 stars in the Keck/HIRES C02~sample
and, in most cases, have multiple measurements separated by up to one
year.

Throughout this paper, we determine dispersions using a Monte Carlo
Markov Chain (MCMC) method \citep{lane10a, gregory05a} assuming
uniform priors on the velocity and velocity dispersion.  This method
produces the same results as the likelihood maximization technique
described in \citet{walker06a} when the error distributions are
Gaussian.  In the regime where the velocity dispersion approaches the
observational errors on this quantity, the MCMC method allows us to
calculate confidence intervals directly from posterior distributions
without the assumption of Gaussianity.  We confirmed that our MCMC
algorithm reproduces the same dispersion and errors reported by C02
for their HIRES dataset.

Using a single epoch of velocities and the 21 star sample defined in
C02, we measured a Keck/DEIMOS velocity dispersion of $2.5\pm
0.7$\kms, in good agreement with the Keck/HIRES measurement.  We next
combined our 47 independent measurements of these 21 stars (six stars
have only a single measured velocity, while 3 stars have four or more
epochs), and redetermine the velocity dispersion.  Combining our
repeat measurements, we find a lower velocity dispersion of $1.6 \pm
0.7$ \kms, suggesting that unresolved binary stars are indeed
inflating the single epoch dispersion, although not to a level
consistent with the stellar mass alone.

\begin{figure}[t!]
\epsscale{1.1}
\plotone{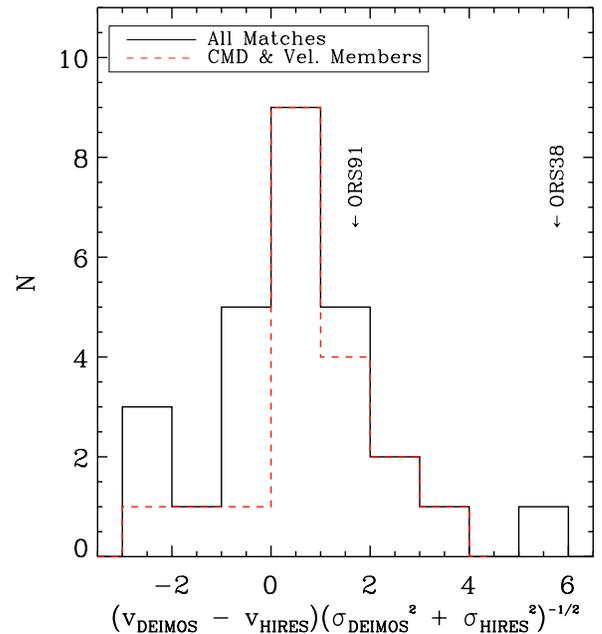}
\caption{Data comparison to Cot\'e \etal HIRES
 velocities. We plot the histogram for the velocity difference,
 normalized by observational error: $(v_{\rm DEIMOS} - v_{\rm
   HIRES})(\sigma_{\rm DEIMOS}^2 + \sigma_{\rm HIRES}^2)^{-1/2}$.
 The two Cot\'e \etal objects in question are identified in their
 respective bins (ORS-91 and ORS-38).  \label{fig_hist_hist}}
\end{figure}

We next examined the C02~stars for evidence of variability.  There are
two stars in this sample which we suspect, ORS-38 and ORS-91,
identified as orange symbols in Figure~\ref{fig_cmd_spat}.  ORS-38 is
near the RGB and varies in velocity by 17\,\kms\ between the DEIMOS
and HIRES measurements.  This is one of the five variables identified
in \S\,\ref{sec_bin_pop}.  While it does not inflate the HIRES
velocity dispersion, ORS-38 is the primary star inflating our
calculations.  ORS-91 is on the horizontal branch and lies in the
instability band identified in \citet{ivezic05a}.  Although the DEIMOS
and HIRES velocities agree, and including/removing this stars does not
affect our results, we chose to exclude this and other horizontal
branch stars from our sample.  Removing ORS-38 and ORS-91 does not
affect the HIRES-determined dispersion, but the DEIMOS-based velocity
dispersion falls to $0.7^{+0.8}_{-0.6}$ \kms.  This dispersion is consistent
with a normal stellar mass-to-light ratio, although our measurement
uncertainties are too large to rule out a higher than expected
mass-to-light ratio.  We plot the error-normalized differences in
velocity between the HIRES and DEIMOS samples in
Figure~\ref{fig_hist_hist}, noting the position of these two rejected
stars.
%Note, the instability band should be 0.11 < g -r < 0.31.  After
%de-reddening,  

\subsection{The Velocity Dispersion of Palomar 13} \label{sec_vel_disp}

We determine the velocity dispersion of Pal\,13 first using the
cleaned sample of 66 stars defined above.  This sample has minimal
Milky Way foreground contamination and contains 5 known velocity variable stars.
Using only a single epoch of DEIMOS observations, we calculate a
dispersion of $2.3\pm0.5$\kms.  Using a weighted average of all our
repeat measurements, the velocity dispersion decreases to
$1.6\pm0.5$\kms.  By removing the 5 known variable stars, the velocity
dispersion drops close to our measurable limits:
$\sigma=0.6^{+0.7}_{-0.5}$\kms.  Finally, adding in quadrature
measurements from C02 where available, the velocity dispersion of
Pal\,13 is $\sigma=0.4^{+0.5}_{-0.4}$\kms.  These results demonstrate
that unresolved binaries or variables, even multiply measured and averaged, can
still contribute significantly to the dispersion given our sample
size.

We take as the final velocity dispersion, the value determine without
binaries combining both DEIMOS and HIRES measurements (61 stars,
$\sigma=0.4^{+0.5}_{-0.4}$\kms).  The average cluster distance of this
sample is $1.5'$, slightly larger than the Pal\,13 half-light radius
of $r_{1/2}=1.3'$, as seen in right panel of
Figure~\ref{fig_vel_vvsr}.  We do not have a sufficient number of stars
to produce a binned velocity dispersion as a function of radius,
however, splitting the sample at the radius enclosing half the
measured stars, we find that both the inner and outer bin have similar
velocity dispersions, but the error on each value doubles.

\subsection{A Stellar Mass-To-Light Ratio} \label{sec_mass2l}

We calculate the mass and mass-to-light ratio of Pal\,13 using the
velocity dispersion determined via our 61 member clean sample without
binary stars, $\sigma =0.4^{+0.5}_{-0.4}$\kms.  We determine the
dynamical mass enclosed within the half-light radius applying the
formula from \citet{wolf10a}: $M_{1/2}(\rm r_{\rm 1/2}) \simeq
4G^{-1}\langle \sigma_{\rm los}^2 \rangle \rm r_{\rm 1/2}$, where
$M_{1/2}$ is the mass contained within the 3D projected half-light
radius, $\sigma_{\rm los}$ is the line-of-sight velocity dispersion.
This mass estimator is based on the Jeans Equation and is less
sensitive to uncertainties in the velocity anisotropy than other
simple mass estimates.  We calculated an enclosed mass of $M_{1/2} <
1.3^{+2.7}_{-1.3} \times 10^3 \Msun$.  Using the luminosity determined
in \S\,\ref{subsec_struct},
$L_{V}=1.1^{+0.5}_{-0.3}\times10^{3}$\,L$_{\sun}$, we calculated a
mass-to-light ratio of $\Upsilon_V =2.4^{+5.0}_{-2.4} \Msun/\Lsun$
within the half-light radius.

We alternatively calculate the mass-to-light ratio of
Pal\,13 assuming velocity isotropy and that the cluster is well fit by
a King profile.  Following C02, the mass-to-light ratio is given by:
$\Upsilon_V = \sigma_0^2/(0.003 r_c \mu_{0,V})$ \citep{richstone86a},
where $r_c$ is the King core radius in parsecs, $\mu_{0,V}$ the
central surface brightness ($\Lsun$\,pc$^{-2}$) and $\sigma_0$ the
central velocity dispersion.  Using the best-fitting King parameters
from \S\,\ref{subsec_struct},  we determine $\Upsilon_V = 1.9$. 

Within the limits of our velocity precision, the mass-to-light ratio
of Pal\,13 is consistent with its stellar mass.  Our estimates are
consistent with the average dynamical mass-to-light ratio of old Milky
Way globular clusters of $\Upsilon_V =1.45$ \citep{mclaughlin00a}.
Since Pal\,13 is slightly more metal-poor than the average Milky Way
globular cluster in this sample, we computed the theoretical
mass-to-light ratio, based on a Chabrier initial mass function, an age
of 12\,Gyr and [Fe/H] $=-1.6$ \citep{bruzual03b} to be $\Upsilon_V
=2.0$.  We therefore conclude that the dynamics of Pal\,13 are
consistent with its stellar mass alone.

\subsection{Spectroscopic Metallicity}\label{subsec_metal}

\begin{figure}[t!]
\epsscale{1.0}
\plotone{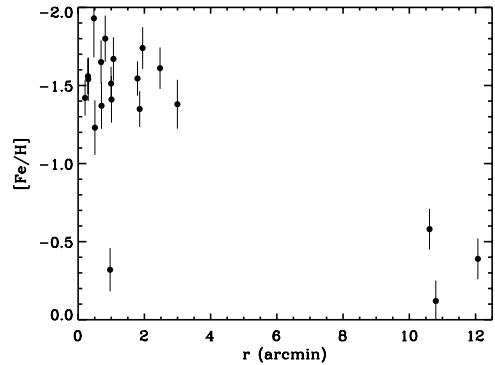}
\caption{Spectroscopic metallicity as a function of cluster distance.
  We reject as Pal\,13 members the four stars with metal-rich
  measurements.  The average metallicity of Pal\,13 is [Fe/H]
  $=-1.6\pm 0.1$, with an internal metallicity spread
  $0.1\pm0.1$\,dex. \label{fig_metals}}
\end{figure}

We measured spectroscopic metallicities for individual stars in
Pal\,13 using the spectral synthesis method introduced by
\citet{kirby08a} and refined by \citet{kirby10a}.  Metallicities based
on this method are reliable for stars with log\,$g < 3.6$ (at the
distance of Pal\,13 roughly $r < 20$).  In Figure~\ref{fig_metals}, we
plot metallicities of the 20 stars passing this criterion in our full
sample.  As discussed in \S\,\ref{sec_membership}, four stars have
clearly deviant metallicities from the main sample and we have excluded
these stars from Pal\,13 membership.  The weighted average metallicity
for the 16 member stars is $\langle$[Fe/H]$\rangle$ = $-1.6 \pm
0.1$\,dex.  This is comparable to previous spectroscopic estimates of
the metallicity based on medium resolution spectroscopy, [Fe/H] =
$-1.67\pm 0.15$ \citep{zinn82a} and [Fe/H] = $-1.9\pm 0.1$,
\citep{friel82a}, as well as [Fe/H] = $-1.98\pm 0.31$\,dex based on a
single Pal\,13 star observed at high resolution \citep{cote02a}.

Using the same sample of 16 stars above, we estimated an internal
metallicity dispersion of $0.1\pm0.1$\,dex.  While the one-sigma
errors are consistent with no internal metallicity dispersion, as
expected for a single stellar population globular cluster, the
measurement does allow a modest metallicity dispersion.  This
dispersion is smaller than observed in some luminous globular clusters
\citep[e.g., NGC~2419][]{cohen10a}, and is significantly smaller than
observed for stellar systems containing dark matter, e.g.~dwarf
galaxies, show clear metallicity dispersions on the order of 0.5\,dex
\citep{kirby10a}.  Dwarf galaxies also exhibit a tight correlation
between luminosity and metallicity \citep{kirby08b}.  Given Pal\,13's
low luminosity ($M_V = -2.8$), this relationship predicts a
metallicity of [Fe/H]$\sim-3$, significantly more metal-poor than
observed.  Both the average metallicity of Pal\,13 and the low
internal metallicity dispersion are further evidence that Pal\,13 is a
globular cluster.

\subsection{Blue Straggler Stars}\label{sec_blues}

We show above that the dynamics of Pal\,13 are influenced by the
presence of unresolved binary stars.  This is unlike other Milky Way
globular clusters whose single-epoch velocity dispersions are
consistent with their stellar mass \citep{baumgardt09a, lane10a,
  hankey10a}.  We next investigate the blue straggler stars (BSSs)
population of Pal\,13 as a means to determine whether the binary
fraction in Pal\,13 is indeed higher than typical globular clusters.
BSSs are positioned brighter and blueward of the primary main-sequence
turnoff.  In globular clusters, BSSs mimic a younger stellar population, but
are interpreted as systems which formed via stellar collisions
\citep{hills76a} or mass transfer in close binary systems
\citep{mccrea64}.  We have not included BSSs in our kinematic
analysis, however, our spectroscopic observations of these stars
provide some clue to their formation, and their frequency provides
further evidence for an enhanced binary fraction in Pal\,13.

In our spectroscopic sample, 12 stars pass the BSS criterion in
color-magnitude (Figure~\ref{fig_cmd_spat}).  Two of these stars do
not pass our velocity criterion for membership, and their proper
motions measured by \citet{siegel01a} confirm they are not Pal\,13
members.  The remaining ten stars have proper motions consistent with
Pal\,13 membership.  We obtained repeated velocity measurements for
only one member star; the repeat measurements are consistent with a
constant velocity.  However, the velocity distribution of the 10 stars
identified as BSS members of Pal\,13 is broader than the overall
velocity distribution of Pal\,13.  This is reflected in the velocity
dispersion of the BSSs, $4.1\pm1.4$\kms\, substantially larger than
the main Pal\,13 sample.  This hints that at least some of these BSSs
may be close binary systems.

A common way to express the BSS abundance is by measuring the ratio,
of BSSs to RGB stars, $F{}_{rgb}^{bss}$. Using our CFHT photometric
data, we estimated the frequency ($F{}_{rgb}^{bss}$) of BSSs following
the method of \citet{leigh07a} by counting the number of BSSs and RGB
stars within well defined regions of the CMD.  These regions are
similar to Leigh \& Sills, with the main difference being our BSS box
is located closer to the main sequence turnoff.  The reason for this
is that Pal\,13 is a low density cluster and overcrowding near the
center is not an issue and, therefore, the scatter in the main
sequence is lower.

We measured $F{}_{rgb}^{bss}$ in both a region within the half-light
radius ($1.3'$) and three times this value obtaining $F{}_{rgb}^{bss}
= 1.0\pm0.4$ and $0.8\pm0.3$, respectively.  Our results are directly
comparable to the values of $F{}_{rgb}^{bss}$ determined by
\citet{leigh07a} for 57 Milky Way globular clusters (their sample did
not include Pal\,13).  These authors found an average value of
$F{}_{rgb}^{bss} = 0.14$ with a standard deviation of 0.19.  In
comparison, Pal\,13 presents a remarkably high frequency of BSSs.
Given the binary fraction measured by \citet{clark04a} of $30\pm4$\%
based on stars redwards of the main sequence, the high frequency of
BSSs can be interpreted as due to mass transfer between primordial
binaries.  The high BSS frequency is consistent with Pal\,13 having a
higher binary fraction than typical Milky Way globulars.

\section{Discussion and Conclusions} \label{sec_discussion}

We demonstrate that the mass-to-light ratio of the Milky Way globular
cluster Pal\,13 is consistent with its stellar mass, based on deep
CFHT/MegaCam imaging and Keck/DEIMOS spectroscopy.  Our spectroscopic
member sample is triple that of previous studies, consisting of 61
stars with multiply measured velocities on the timescale of days and
years.  We have minimized contamination from Milky Way foreground
stars via proper motion estimates from \citet{siegel01a}, and removed
unresolved binary stars when possible via multiple spectroscopic
measurements.  When combined with HIRES velocities from C02, our final
velocity dispersion for Pal\,13 is $\sigma = 0.4 ^{+0.4}_{-0.3}$\kms,
corresponding to a mass-to-light ratio of $\Upsilon_V = 2.4
^{+5.0}_{-2.4} \Msun/\Lsun$.  Thus, the dynamical mass of Pal\,13 is
consistent with its stellar mass and there is no need for dark matter.
We therefore confirm the suggestion by \citet{blecha04a} that binaries
inflate the single epoch velocity dispersion.  While
Pal\,13 appears to be a 'normal' globular cluster in this sense, two
independent observed quantities in Pal\,13 are atypical: (1) unlike
other Milky Way globular clusters, the single epoch velocity
dispersion is higher than predicted based on the stellar mass alone,
and (2) the surface brightness slope is shallower than typical
globular clusters.  We examine each of these statements below.

The single epoch velocity dispersion of Pal\,13 is inflated due to
unresolved binary stars in both the HIRES data from C02 and our own
DEIMOS dataset.  We arrive at our final dispersion by removing these
binary systems.  Yet unresolved binaries do not affect the velocity
dispersion of other Milky Way globular clusters
\citep{baumgardt09a,lane10a}, at least not to the same extent.  Thus,
Pal\,13 must have a higher binary fraction as compared to these
globular clusters.  Estimates from the main sequence \citep{clark04a}
and blue straggler stars (\S,\ref{sec_blues}; Santana et al., ~2012,~in prep)
are consistent with this statement.  However, we are necessarily
comparing Pal\,13's binary fraction to luminous and more concentrated
globular clusters where such estimates exist. 

A stellar mass-to-light ratio for Pal\,13 does not provide an
explanation for the second anomalous observations in Pal\,13, the
observed shallow surface brightness ($\eta = -2.8\pm0.3$), nor the
large King tidal radius ($r_t = 98\pm 11$\,pc) which is significantly
larger than the calculated Jacobi tidal radius of 47\,pc.  The
half-light radius of Pal\,13 ($r_{1/2} = 1.27\pm0.16\arcmin =
9.0\pm1.1$\,pc) is larger than the average Milky Way cluster, as well
as the median size of globular clusters in external galaxies
\citep[e.g.,][]{jordan09a, masters10a}.  Pal\,13 may be a Galactic
analog of the "diffuse star clusters" seen in some other galaxies
\citep{brodie02a, Peng06a}.  

The photometric isophotes of Pal\,13 (Figure~\ref{fig_profile}) show
some irregularity near the Jacobi radius, but we do not observe the
well defined S-shape curve seen in other globular clusters which are
actively being tidally stripped.  While Pal\,13 may not currently be
undergoing tidal heating (e.g.~lack of S-curve in photometry, low
velocity dispersion), tidal debris in the vicinity of the cluster may
influence its appearance.  \citet{kupper11a} proposed that Pal\,13 is
at apogalacticon (farthest approach), and the photometric properties
can be explained by tidal debris stripped from the cluster throughout
its orbit and compressed.  Confirming whether the photometric features
of Pal\,13 are intrinsic or due to tidal interactions may be tested by
future studies that consist of measuring velocities and metallicities
of stars at larger radius then the present sample, particularly
between the predicted and King tidal radius ($6-13'$) of Pal\,13.

While the large inferred radii of Pal\,13 may be due to tidal
interactions, its present luminosity and size are currently more
closely related to the dark matter dominated Milky Way ultra-faint
galaxies (Figure~\ref{fig_sizeMV}).  Is it possible that the
ultra-faint galaxies are also affected by tidal interactions or binary
stars, and that these objects may instead have mass-to-light more
similar to globular clusters \citep[e.g.,][]{mcConnachie10a}?  This
exercise has been done for the ultra-faint galaxies Segue\,1
\citep{geha09a,simon11a} and Bootes\,I \citep{koposov11a}, whose
velocities dispersions are between 2-4\kms.  In both
cases, the binary-corrected velocity dispersions are still consistent
with a significant dark matter content.  Furthermore, both galaxies
have a more significant spread in metallicities as compared to
Pal\,13.    Thus, both the corrected velocity dispersions and
metallicities suggest that the ultra-faint galaxies are distinct from
low luminosity globular clusters such as Pal\,13.

The dynamical mass of Pal\,13 inferred from the binary-corrected
velocity dispersion is consistent with its stellar mass alone.
However, the large influence of unresolved binaries on the velocity
dispersion, as well as high inferred blue straggler population makes
Pal\,13 a unique system.  In addition Pal\,13 has an unusual density
structure suggesting that tidal interaction may be important in this
low-mass halo object.  Studying other low luminosity globular clusters
in the Milky Way halo and beyond will answer whether these properties
are unique to Pal\,13 or are typical properties of low luminosity globular clusters.\\

JDB acknowledges support from the CT Space Grant.  MG acknowledges
support from NSF grant AST-0908752 and the Alfred P.~Sloan
Foundation. RRM acknowledges support from the GEMINI-CONICYT Fund,
allocated to the project N°32080010 and from CONICYT through projects
FONDAP N°15010003 and BASAL PFB-06.  Support for this work was
provided by NASA through Hubble Fellowship grant 51256.01 awarded to
ENK by the Space Telescope Science Institute, which is operated by the
Association of Universities for Research in Astronomy, Inc., for NASA,
under contract NAS 5-26555. SGD acknowledges a partial support from
the NSF grant AST-0909182.  We would like to thank Andreas
K\"{u}epper and Luis Vargas for useful conversation.

\newpage

\begin{deluxetable}{lccrrccc} \label{table_mask}
\tabletypesize{\scriptsize}
\tablecaption{Keck/DEIMOS Multi-Slitmask Observing Parameters}
\tablewidth{0pt}
\tablehead{
\colhead{Mask} &
\colhead{$\alpha$ (J2000)} &
\colhead{$\delta$ (J2000)} &
\colhead{Date Observed}&
\colhead{PA} &
\colhead{$t_{\rm exp}$} &
\colhead{\# of slits} &
\colhead{\% useful} \\
\colhead{Name}&
\colhead{(h$\,$:$\,$m$\,$:$\,$s)} &
\colhead{($^\circ\,$:$\,'\,$:$\,''$)} &
\colhead{}&
\colhead{(deg)} &
\colhead{(sec)} &
\colhead{}&
\colhead{spectra}
}
\startdata
Pal13\_1 & 23:07:03.4  &  +12:50:55  & Sept 4, 2008 & 38        & 9000 &  74 & 86\%\\
Pal13\_2 & 23:06:50.8  &  +12:49:09  & Sept 4, 2008 & $-$13  & 7200 &  81 & 95\%\\
Pal13\_3 & 23:06:47.3  &  +12:44:14  & Sept 5, 2008 & 112      & 7200 &  72 & 94\%\\
Pal13\_4 & 23:06:41.7  &  +12:43:45  & Sept 5, 2008 & 180      & 8100 &  83 & 95\%\\
Pal13\_5 & 23:06:08.4  &  +12:49:16  & Sept 4, 2008 & 17        & 2700 &  52 & 96\%\\
Pal13\_6 & 23:07:22.7  &  +12:49:25  & Sept 4, 2008 & $-$166& 2460 &  56 & 98\% \\
Pal13\_7 & 23:06:31.2  &  +12:37:44  & Sept 5, 2008 & 87        & 2400 &  51 & 92\%\\
Pal13\_8 & 23:06:42.3  &  +12:45:54  & Sept 5, 2008 & 74        & 1680 &  42 & 83\%\\
pal13    & 23:06:51.0    &  +12:45:14  & Oct 13, 2009 & 0         & 1800 &  23 & 96\%
\enddata
\tablecomments{The right ascension, declination, date of observation, 
  position angle and total exposure time for each Keck/DEIMOS
  slitmask.  The final two columns refer to the total number of
  slitlets on each mask and the percentage of those slitlets for which
  a redshift was measured.}
\end{deluxetable}

\begin{deluxetable}{lll}
\label{table_prop}
\tabletypesize{\scriptsize}
\tablecaption{Palomar 13 properties}
\tablewidth{0pt}
\tablehead{
\colhead{Property}&
\colhead{\Cote~\etal (2002)}&
\colhead{This work}
}
\startdata
Core Radius (King Model), $r_c$& $0.23 \pm 0.03 $ arcmin & $0.42 \pm 0.06$ arcmin \\
Tidal Radius (King Model), $r_t$& $26 \pm 6$ arcmin & $13.9 \pm 1.50$ arcmin \\
Half-Light Radius (King Model), $r_{1/2}$ & N/A & $1.27 \pm 0.16$ arcmin \\
Luminosity, $L_{V}$ & $(2.8 \pm 0.4) \times 10^3 \Lsun$ & $(1.1^{+0.5}_{-0.3}) \times 10^3 \Lsun$ \\
Absolute Magnitude, $M_{V}$ & $-3.8$ mag & $-2.8 \pm 0.4$ mag \\
Metallicity, $\rm [Fe / \rm H]$ & $-1.9 \pm 0.2$ dex & $-1.6 \pm 0.1$ dex \\
Right Ascension (Center, King Model), $ra$ & $346.6867$ deg & $346.68519 \pm 0.00063$ deg \\
Declination (Center, King Model), $dec$ & $12.7717$ deg & $12.771539 \pm 0.00068$ deg \\
Distance Modulus, $(m-M)_0$ & $16.93 \pm 0.10$ mag & N/A \\
Distance, $d$ & $24.3^{+1.2}_{-1.1} $ kpc & N/A \\
\enddata
\tablecomments{We list properties of Pal 13 compiled by C02 and as independently
  measured by this work for comparison.}
\end{deluxetable}

\begin{deluxetable}{ccccccccccc}
\tabletypesize{\scriptsize}
\tablecaption{DEIMOS and HIRES Velocity Measurements for Palomar 13}
\tablewidth{0pt}
\tablehead{
\colhead{i} &
\colhead{Name} &
\colhead{$\alpha$ (J2000)} &
\colhead{$\delta$ (J2000)} &
\colhead{$r-$mag} &
\colhead{$(g-r)$} &
\colhead{$v_{\rm avg}$} &
\colhead{$v_{\rm D}$} &
\colhead{$v_{\rm H}$} &
\colhead{Full} &
\colhead{Clean} \\
\colhead{}&
\colhead{}&
\colhead{(h$\,$ $\,$ m$\,$ $\,$s)} &
\colhead{($^\circ\,$ $\,'\,$ $\,''$)} &
\colhead{(mag)} &
\colhead{(mag)} &
\colhead{(\kms)} &
\colhead{(\kms)} &
\colhead{(\kms)} &
\colhead{}  &
\colhead{} 
}
\startdata
  0 & LRIS\_61 & 23:06:43.15 & +12:44:50.30 & 19.85 &  0.44 &   29.85 $\pm$ 2.85 &   29.85 $\pm$ 2.85 & & N & N \\
  1 & LRIS\_21 & 23:06:42.00 & +12:45:26.50 & 17.68 &  0.75 &   20.27 $\pm$ 0.26 &   27.13 $\pm$ 2.25 &   20.18 $\pm$ 0.26 & Y & Y \\
  2 & LRIS\_47 & 23:06:45.59 & +12:45:39.50 & 19.36 &  0.68 &   23.37 $\pm$ 0.69 &   35.15 $\pm$ 2.23 &   22.14 $\pm$ 0.72 & Y & Y \\
  3 & LRIS\_22 & 23:06:48.27 & +12:45:46.60 & 18.30 &  0.61 &   19.98 $\pm$ 0.46 &   25.27 $\pm$ 2.21 &   19.74 $\pm$ 0.47 & Y & Y \\
 & & & & & & &   19.61 $\pm$ 2.23 &   19.93 $\pm$ 0.54 & &  \\
 & & & & & & &   29.36 $\pm$ 2.22 &   19.41 $\pm$ 0.93 & &  \\
 & & & & & & &   25.72 $\pm$ 2.22 &   18.87 $\pm$ 1.84 & &  \\
 & & & & & & &   21.50 $\pm$ 2.27 & & & \\
  4 & LRIS\_64 & 23:06:41.61 & +12:46:09.50 & 19.91 &  0.54 &   30.42 $\pm$ 2.61 &   30.42 $\pm$ 2.61 & & Y & Y \\
 & & & & & & &   32.39 $\pm$ 2.85 & & & \\
 & & & & & & &   27.43 $\pm$ 3.13 & & & \\
  5 & LRIS\_41 & 23:06:44.76 & +12:46:18.80 & 19.75 & -0.12 &   25.94 $\pm$ 3.41 &   25.94 $\pm$ 3.41 & & N & N \\
 & & & & & & &   25.94 $\pm$ 3.41 & & & \\
  6 & LRIS\_15 & 23:06:48.52 & +12:46:19.20 & 17.32 &  0.72 &   29.22 $\pm$ 0.27 &   24.68 $\pm$ 2.20 &   29.29 $\pm$ 0.27 & Y & Y \\
 & & & & & & &   23.92 $\pm$ 2.21 &   29.14 $\pm$ 0.80 & &  \\
 & & & & & & &   25.01 $\pm$ 2.20 &   30.55 $\pm$ 0.64 & &  \\
 & & & & & & &   24.65 $\pm$ 2.22 &   28.76 $\pm$ 0.36 & &  \\
 & & & & & & & &   29.80 $\pm$ 0.62 & & \\
  7 & LRIS\_39 & 23:06:43.96 & +12:46:20.20 & 19.33 &  0.58 &   25.60 $\pm$ 0.68 &   30.15 $\pm$ 2.38 &   25.19 $\pm$ 0.71 & Y & N \\
 & & & & & & &   30.15 $\pm$ 2.38 &   25.19 $\pm$ 0.71 & &  \\
 ... & ... & ... & ... & ... & ... & ... & ... & ... & ...
\enddata
\tablecomments{Velocity measurements, magnitude, color and membership
  data for spectroscopically observed stars of Pal\,13.  Position,
  apparent $r-$band magnitude, $(g-r)$ color, weighted average of
  DEIMOS and HIRES heliocentric velocities ($v_{\rm avg}$), DEIMOS heliocentric
  radial velocity ($v_{\rm D}$),HIRES heliocentric radial velocity
  ($v_{\rm H}$) from C02, member and clean member status for each star
  as determined in \S\,\ref{sec_data} and \S,\ref{sec_membership}.
  Each sub-row lists the subsequent DEIMOS and HIRES individual epoch
  velocity measurements.  Note that the HIRES velocities have been
  shifted by $+ 0.5$ \kms for zero-point alignment to this study.
  This table will be published in its entirety in the electronic
  edition of the {\it Astrophysical Journal}.  }
\end{deluxetable}

\end{document}